\documentclass[fleqn,usenatbib]{mnras}

\usepackage[T1]{fontenc}

\DeclareRobustCommand{\VAN}[3]{#2}
\let\VANthebibliography\thebibliography
\def\thebibliography{\DeclareRobustCommand{\VAN}[3]{##3}\VANthebibliography}

\usepackage{graphicx}	
\usepackage{amsmath}	
\usepackage{amssymb}	
\usepackage[normalem]{ulem}

\title[Transient hot and cool loops]{Spectroscopic and imaging observations of transient hot and cool loops by IRIS and SDO}

\author[Gupta \& Nayak]{Girjesh R. Gupta$^{1}$\thanks{E-mail: girjesh@prl.res.in (GRG)},
Sushree S. Nayak$^{1,2}$
\\
$^{1}$ Udaipur Solar Observatory, Physical Research Laboratory, Dewali, Badi Road, Udaipur 313001, India \\
$^{2}$ Department of Physics, Indian Institute of Technology Gandhinagar, Palaj, Gandhinagar 382355, India \\
}

\date{Accepted 2022 March 04. Received 2022 February 14; in original form 2021 July 07}

\pubyear{2022}

\begin{document}
\label{firstpage}
\pagerange{\pageref{firstpage}--\pageref{lastpage}}
\maketitle

\begin{abstract}
Coronal loops are basic building blocks of solar atmosphere and are observed on various length scales. However, their formation mechanism is still unclear. In this paper, we present the spectroscopic and imaging observations of small-scale transients and subsequent formation of transient loops. For the purpose, we have utilized the multi-wavelength observations recorded by Atmospheric Imaging Assembly   (AIA) and Interface Region Imaging Spectrometer (IRIS) Slit-Jaw-Imager (SJI), along with spectroscopic measurements provided by IRIS. For the photospheric magnetic field data, we obtained line-of-sight magnetogram data provided by Helioseismic and Magnetic Imager (HMI). Small-scale transients are simultaneously observed with several EUV and UV passbands of AIA and IRIS-SJI. HMI magnetogram provides evidence of negative flux cancellations beneath these transients. Differential Emission Measure (DEM) analysis shows that one of the transient attains temperature up to 8 MK whereas another one reaches only up to 0.4 MK. These transients further lead to the formation of small-scale loops with similar temperature distributions, and thus termed as hot and cool loops respectively. During the course of events, IRIS slit was rastering the region and thus provided spectroscopic measurements at both transients and associated loops. This enabled us to perform in-depth investigations of hot and cool loops. Using density sensitive O IV line pair, we obtained average electron densities along the hot and cool loop to be $10^{11.2}$ and $10^{10.8}$ cm$^{-3}$ respectively. Energy estimates suggest that flux cancellation can easily power the hot transient whereas is insufficient for cool transient. Life time estimates and magnetic field extrapolation suggest presence of small-scale and fine structures within these loops. Results provide crucial ingredients on the physics of loop formation and involved thermodynamics.
\end{abstract}

\begin{keywords}
Sun: corona -- Sun: transition region -- Sun: UV radiation -- Sun: flares -- line: profiles
\end{keywords}


\section{Introduction}
\label{sec:intro}

The solar atmosphere is highly dynamic and is structured by the presence of different type of loops of various length scales. However, these loops are classified as cool loop, warm loop, hot loop, and flaring loop based on their temperature as <1 MK, 1--2 MK, 2--4 MK, and 8--10 MK respectively \citep[e.g.,][]{2014LRSP...11....4R}. Heating of such loops is still an unsolved problem and fall under the greater problem of coronal heating \citep[for details, see][]{2005psci.book.....A,2006SoPh..234...41K}. Magnetohydrodynamic (MHD) waves and magnetic reconnection are believed to play an important role in the coronal heating, however, contributing mechanisms are yet to be identified \citep[for current status of progress, see][]{2012RSPTA.370.3217P,2015RSPTA.37340269D}.    

\begin{figure*}
	\includegraphics[width=0.4\textwidth]{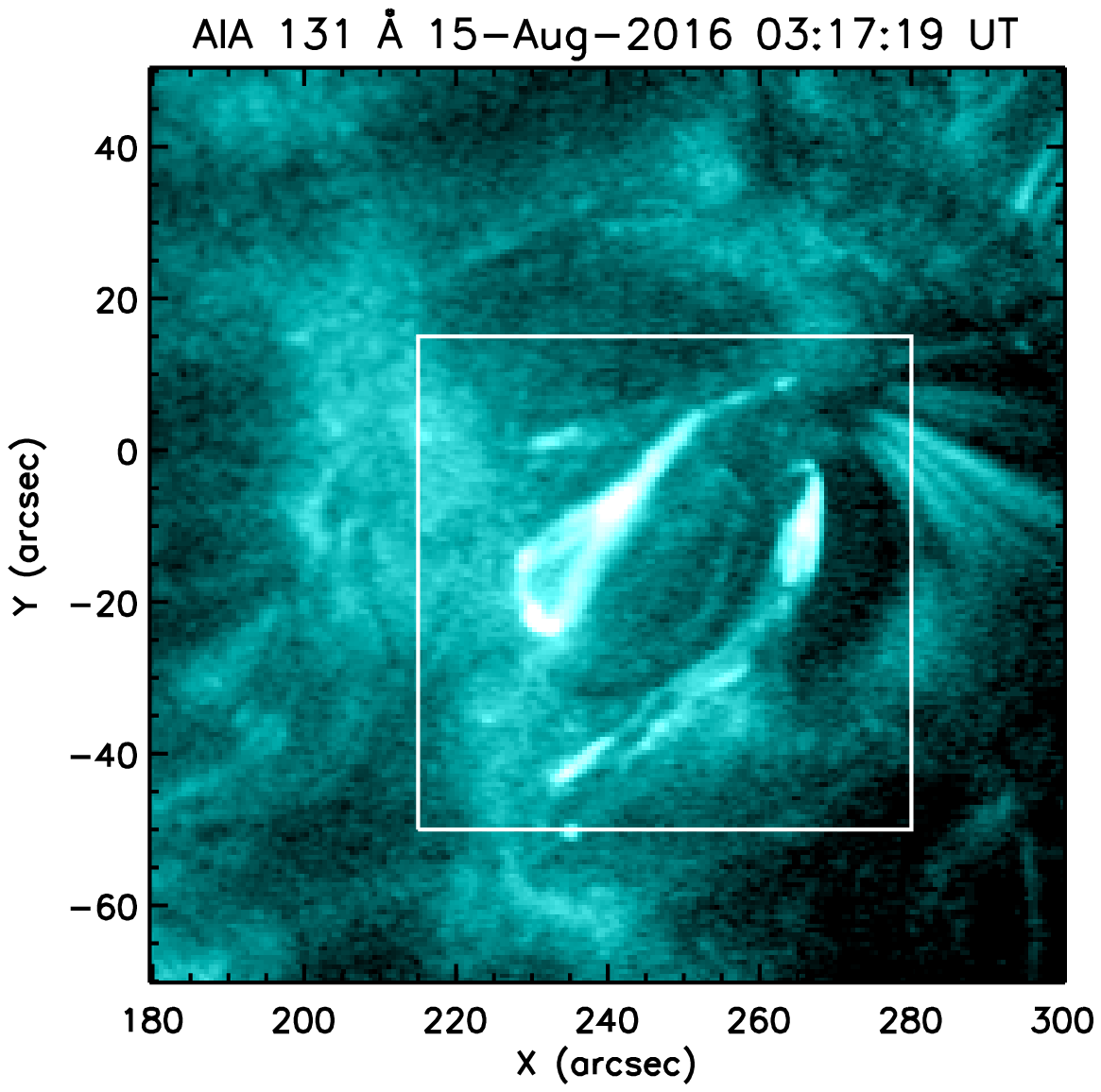}\includegraphics[width=0.6\textwidth]{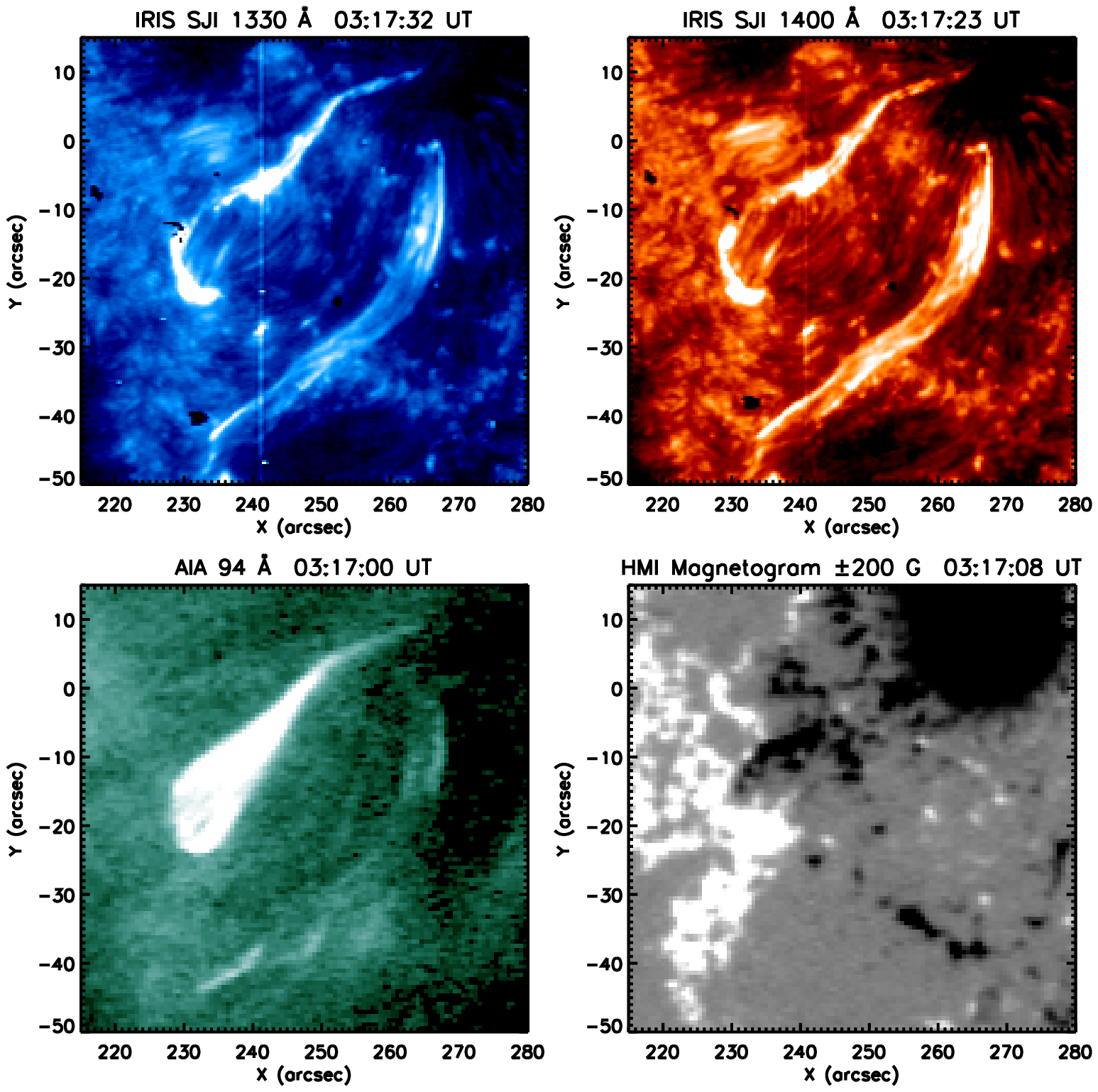}
    \caption{Left panel: AIA 131~{\AA} image recorded on 15 August 2016 shows location of two transient loops. Over-plotted box indicates region chosen for detailed study. Right panels show box region as observed from IRIS-SJI 1400 \AA , AIA 131 and 94 \AA\  passbands, and underlying HMI line-of-sight magnetogram as labeled.}
    \label{fig:context}
\end{figure*}

The solar atmosphere is highly inhomogeneous and hosts several small-scale dynamic structures and events. These small-scale events may provide clues for the mass and energy supply to the upper atmosphere. With the launch of high resolution imaging and spectroscopic instrument Interface Region Imaging Spectrometer \citep[IRIS;][]{2014SoPh..289.2733D} several small-scale transients have been observed with varying properties \citep[see for e.g.,][]{2018SSRv..214..120Y}. Not only these, several group of cool transition region loops are also found in abundant \citep[e.g.,][]{2015ApJ...810...46H}. These loops are low-lying and vary rapidly on the time scales of few minutes \citep[e.g.,][]{2014Sci...346E.315H}. Moreover, several examples of low-lying, small-scale rapidly varying loops on the time scales of few minutes but at flaring loop temperatures also exist in the solar atmosphere \citep[e.g.,][]{2014Sci...346B.315T,2018ApJ...857..137G}. Foot-points of these loops show several brightening activities on various scales and are generally associated with magnetic flux cancellation events at the photosphere \citep[e.g.,][]{2015ApJ...810...46H,2018A&A...615L...9C}. Recently, \citet{2020A&A...644A.130C} found majority of hot coronal loops have at least one foot-point rooted in regions of interacting mixed magnetic polarity and provided spectroscopic evidence for magnetic reconnection at these locations. Therefore, observations of formation of these small-scale loops along with evolution of their magnetic foot-points may provide crucial information on their heating mechanisms \citep{2018ApJ...862L..24P}.

Small-scale loops observed by IRIS do not only show intermittent brightening, but are also associated with excess line broadening \citep{2014Sci...346E.315H}. For some of the loops, wing enhancements of up to 200 km~s$^{-1}$ were observed \citep{2015ApJ...810...46H}. Recently, \citet{2020NatAs.tmp..242B} attributed these enhanced line wings to ion cyclotron turbulence which resulted  due to strong current at reconnection sites. Therefore, spectroscopic observations of such transient loop brightenings can provide important information on reconnection mediated energy release in the solar atmosphere, and thus, in our understanding of solar atmospheric heating. Recently, \citet{2021ApJ...909..105T} and \citet{2021arXiv210503199H} studied the formation of transient loops in the active region using imaging data. They found evidence of magnetic reconnection to be likely source of such transients and loops which were multi-thermal in nature.    

Based on the above observations, it will be important to investigate the nature of magnetic energy sources and topologies that is likely responsible for powering such hot and cool loops. It will also be interesting to investigate nature of different type of energies being released during the loop formations. In the present work, we provide imaging and spectroscopic properties of two nearly simultaneous and nearby transient loops. Among these, one loop achieved flaring loop temperature (hot loop) whereas another reached only up to transition region temperature (cool loop). The paper is organised as follows. In Section~\ref{sec:obs}, we present multi-wavelength observations of two loops and transients. Their properties are described in Section~\ref{sec:analysis}. Magnetic field topology of these loops are described in Section~\ref{sec:nfff}. Finally, we summarize and discuss our results in Section~\ref{sec:discussion}.

\section{Observations}
\label{sec:obs}

\begin{figure*}
	\includegraphics[width=0.9\textwidth]{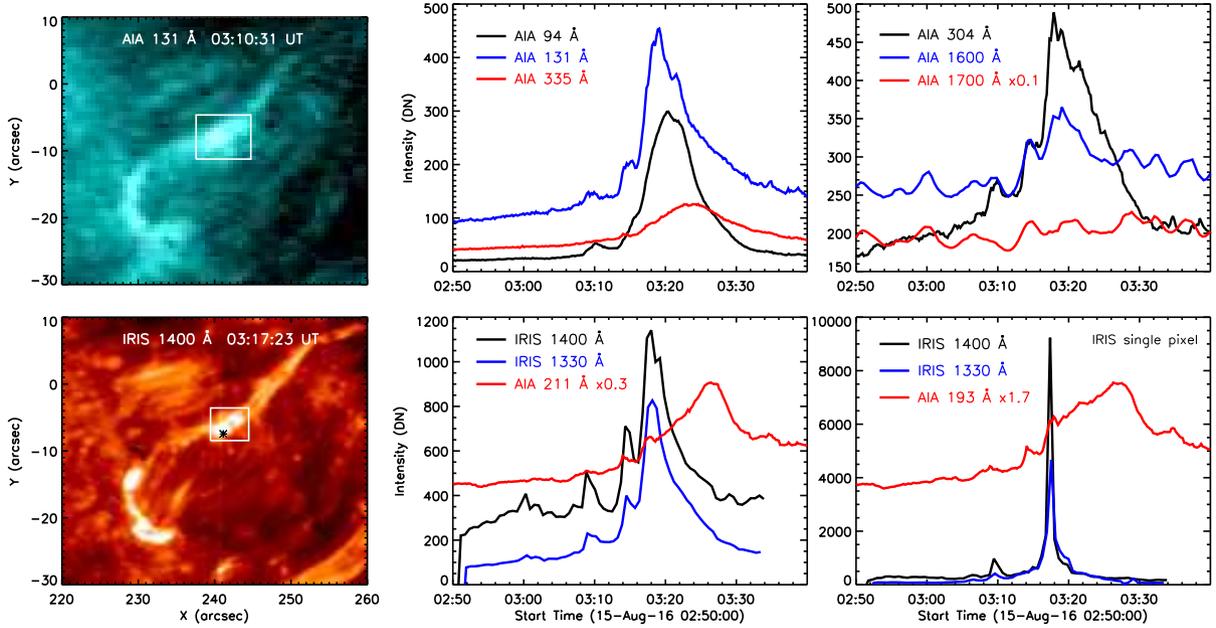}
    \caption{Left panels: Images obtained from AIA 131 \AA , and IRIS-SJI 1400 \AA\  passbands. Overplotted white box marks region of hot transient. Right panels show intensity evolution with time in the boxed region observed from different AIA and IRIS passbands as labeled. IRIS single pixel intensity evolution near the brightening region is also plotted whose location is marked with asterisk (*) on the IRIS 1400 \AA\ image.}
    \label{fig:hbevl}
\end{figure*}

To investigate the origin and characteristics of transient hot and cool loops observed in the active region, we identified a suitable dataset observed by Atmospheric Imaging Assembly \citep[AIA;][]{2012SoPh..275...17L}, Interface Region Imaging Spectrograph \citep[IRIS;][]{2014SoPh..289.2733D}. To trace the magnetic origin of these transient loops in the solar photosphere, we further utilized line-of-sight (LOS) magnetic field data obtained from Helioseismic and Magnetic Imager \citep[HMI;][]{2012SoPh..275..207S,2012SoPh..275..229S}. AIA and HMI are both on-board Solar Dynamics Observatory \citep[SDO;][]{2012SoPh..275....3P}. IRIS dataset is chosen because it shows formation of two nearby transient loops almost at the same time while IRIS spectroscopic slit was rastering the region. IRIS performed large dense rastering of the observed region on 15 August 2016 from 02:45:09 to 03:33:52 UT covering 129'' along the spectral slit with pixel resolution 0.33'' and 320 raster steps in a step size of $\approx 0.37''$. Thus, covered the total field-of-view (FOV) of about $120'' \times 129''$ with exposure time of $\approx 8$ s and effective cadence of $\approx 9$ s. IRIS also observed with Slit-Jaw Imager (SJI) in 1330 and 1400 \AA\ passbands with exposure time of again $\approx 8$ s but with an effective cadence of $\approx 36$ s. IRIS-SJI covered the total FOV of about $234'' \times 129''$. 

Fig.~\ref{fig:context} shows context image of the observations. Left panel shows AIA 131 \AA\ image at a time when both the transient loops were visible almost simultaneously, although at different level of activities. Right panels show zoomed version of two loop regions as visible in IRIS-SJI 1400 \AA , AIA 131 and 94 \AA\  passbands together with underline HMI line-of-sight magnetogram data scaled at $\pm 200$ G. AIA 94 \AA\ image shows presence of only one loop while IRIS 1400 \AA\ image shows clear presence of both loops. All the images shown here are obtained at around 03:17 UT as labeled. IRIS and AIA observations were co-aligned using IRIS-SJI 1400 \AA\ and AIA 1600 \AA\ images with cross-correlation methods. All the AIA, IRIS, and HMI images obtained from different filters were further derotated with respect to time at 02:45:00 UT using the standard Solar Software routines \citep[SSW;][]{1998SoPh..182..497F}. In the study, we utilized IRIS level-2 data.  Slit-jaw images from different filters are already co-aligned. We used IRIS cool neutral line \ion{S}{I} 1401.514~{\AA} to perform the absolute wavelength calibration of IRIS spectral lines. The identified data set provides a unique opportunity to study spatial and temporal evolution of transient loops using both imaging and spectroscopic observations along with the evolution of underlying photospheric magnetic field.

\section{Data Analysis and Results}
\label{sec:analysis}

Two transient loops appeared in the IRIS SJI FOV, however, only one of the loop was visible from AIA 94 \AA\ passband. These bright loops are associated with the several small-scale transients which occurred before the loop brightened up. Some of such compact small-scale transients observed by IRIS are termed as UV burst \citep[described in][]{2018SSRv..214..120Y}. Although, most of the UV bursts are found not to show any significant co-spatial brightening in AIA coronal passbands, however, some weak enhancements were also found sometimes \citep[e.g.,][]{2015ApJ...809...82G}.  However, here we are observing clear intensity enhancements in AIA coronal passbands, thus we will be using the term small-scale transient or compact coronal brightenings for such events \citep[e.g.,][]{2018A&A...615L...9C}. These events are sometimes also associated with brightening of coronal loops \citep[e.g.,][]{2018ApJ...857..137G,2020A&A...644A.130C}. Here we focus our attention on such compact brightenings and associated formation of hot and cool loops. In this section, we describe their imaging, spectroscopic, and magnetic properties in great detail.

\subsection{Imaging Analysis}
\label{sec:imaging}

\begin{figure*}
	\includegraphics[width=0.95\textwidth]{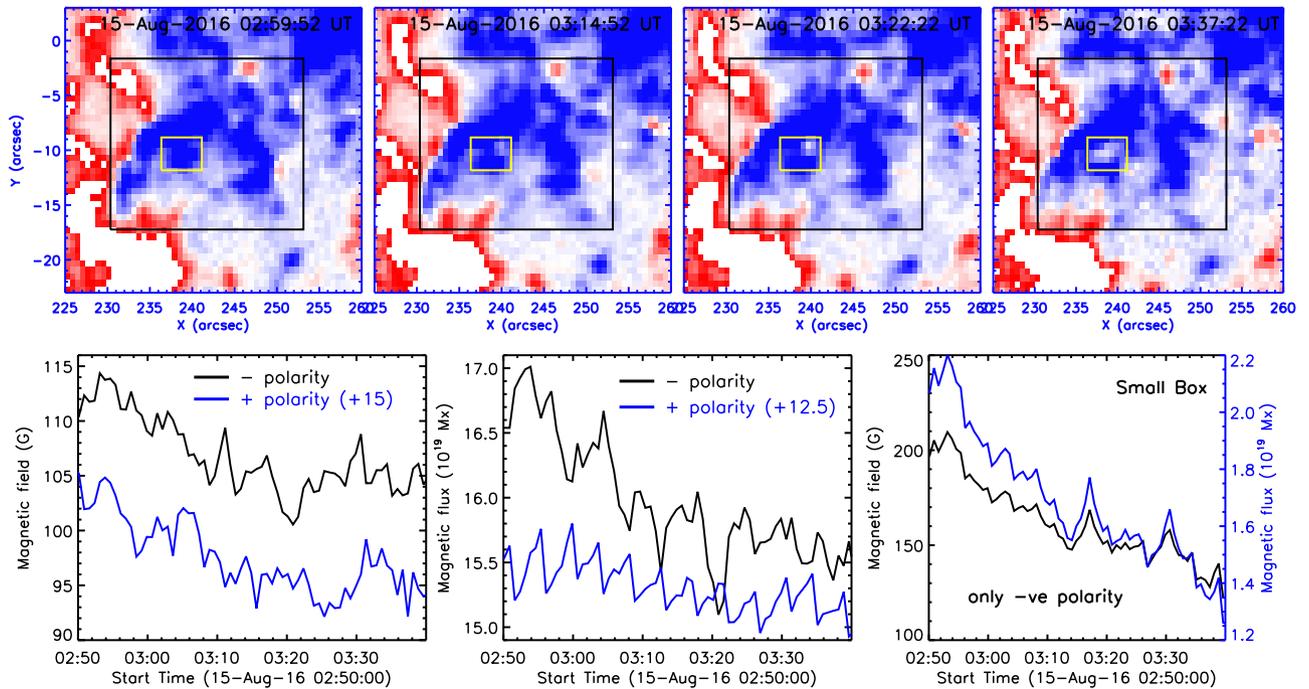}
    \caption{Top panels: images of HMI LOS magnetic field ($\pm 200$ G) beneath the hot transient as identified in Fig.~\ref{fig:hbevl} at different times. Bottom panels: magnetic field and flux evolution within the large and small boxes overplotted in the top panels. Positive polarity data are scaled up to fit in the range.}
    \label{fig:hbmevl}
\end{figure*}

In Fig.~\ref{fig:context}, IRIS-SJI 1330 and 1400 \AA\ images show presence of two very bright loops of length about 30''-40''. 
Upper loop is very clearly visible in the hot coronal filters of AIA such as 94 \AA\ and 131 \AA\ passbands, however, lower loop does not show any significant brightening which were very well seen in IRIS-SJIs. We termed these loops as hot and cool loops respectively. These loops are transient in nature and have life time of few minutes. Hot loop has life time of about 10-11 min whereas cool loop has about 6-7 min. Careful analysis of time evolution of loops from images indicate that these are not a single loop-like structures but collection of several fine-scale threads \citep[e.g.,][]{2014Sci...346E.315H,2015ApJ...810...46H}. However for the simplicity of our analysis, we treat them as a single loop-like structures and study their evolution and properties in detail. Different AIA and IRIS-SJI passbands are sensitive to different plasma temperatures. IRIS-SJI 1330 \AA\ and 1400 \AA\ passbands are sensitive to temperature about 30 kK and 80 kK respectively \citep{2014SoPh..289.2733D}. AIA 131 \AA , 94 \AA , and 335 \AA\ passbands are sensitive to 10 MK, 6 MK, and 2.5 MK plasma temperatures respectively \citep{2012SoPh..275...17L}. 

In Fig.~\ref{fig:hbevl}, we plot time evolution of small-scale brightening event associated with the hot loop. Left panels show location of brightening region (marked with white box) in AIA 131 \AA\ (top) and IRIS-SJI 1400 \AA\ (bottom) filters at different phases of evolution as labeled. Intensity evolution obtained from different AIA and IRIS filters (i.e. at different temperatures) in the boxed region are plotted in right panels as labeled. Evolution of intensity curves indicate that brightening region attained its peak at around 03:18 UT simultaneously in all the cool temperature filters such as AIA 304 \AA , 1600 \AA , and IRIS 1330 \AA and 1400 \AA . However, we did not find any noticeable increase in AIA 1700 \AA\ filter. Interestingly, hot temperature filters such as AIA 131 \AA , 94 \AA , and 335 \AA\ show time dependent appearance of intensity peaks where hottest filter peaks first and followed by cooler ones. Duration of peak width is also temperature dependent where hottest temperature filter shows shortest duration of intensity enhancements. Two filters of AIA 131 \AA , and 94 \AA\ are sensitive to hot plasma representing Fe XXI 128.75 \AA , and Fe XVIII 93.93 \AA\ emissions respectively. However, both of these filters also have significant contributions from cooler temperature plasma \citep{2010A&A...521A..21O,2012ApJ...759..141W,2013A&A...558A..73D}. These cooler components can be removed from AIA 94 \AA\ using the intensities obtained from AIA 211 \AA , and 171 \AA\ passbands. We utilized the method of \citet{2013A&A...558A..73D} and found that emission and enhancements observed at transient location are indeed emitted by hot Fe XVIII 93.93 \AA\ plasma. In the bottom right panel of Fig.~\ref{fig:hbevl}, we also show IRIS light curves obtained at single pixel level. Location of this single pixel is marked with asterisk (*) on the IRIS 1400 \AA\ image. These light curves show brightening enhancement of about $30-40$ times in the both IRIS 1330 and 1400 \AA\ filters. Observations of temperature dependent time evolution of plasmas are commonly found at foot-points of flare loops \citep[e.g.][]{2013ApJ...774...14Q}. Response of foot-point heating is first observed at lower atmosphere as probed by AIA 1600 \AA , IRIS 1330 and 1400 \AA\ and later at higher temperatures which cools down successively as found in AIA 131, 94, 211 \AA\ passbands. Therefore, this hot transient brightening can be categorized as miniature version of standard flare and can also be termed as micro-flare \citep[e.g.][]{2018ApJ...857..137G}.

\begin{figure*}
	\includegraphics[width=0.9\textwidth]{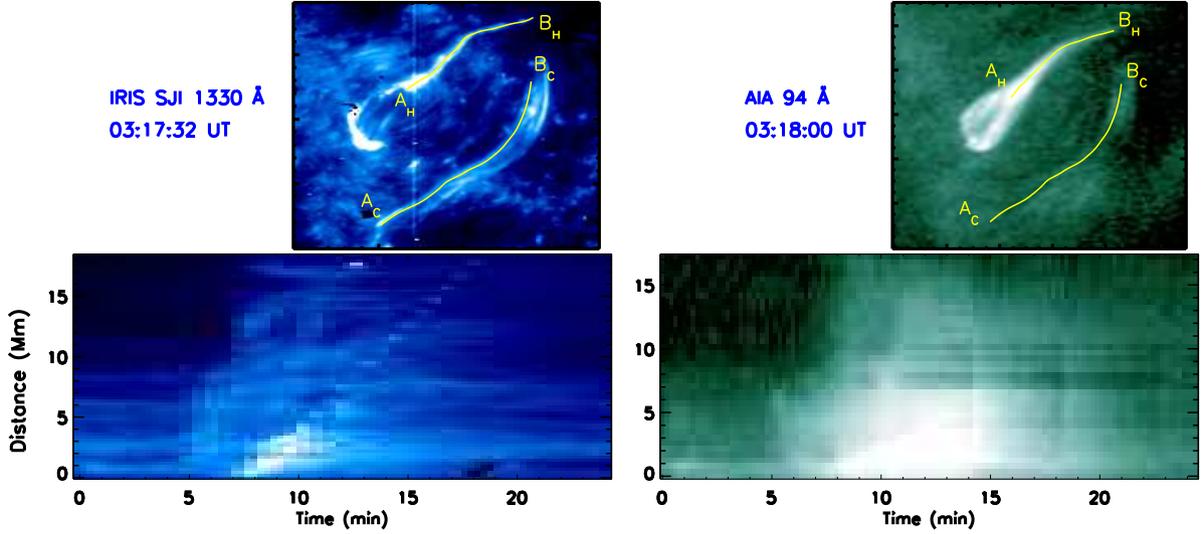}
    \caption{Top panels: images from IRIS-SJI 1330 \AA\ and AIA 94 \AA\ filters showing brightened loops after transients. Loops are traced from points A to B as labeled where subscripts H and C indicate hot and cool loops respectively. Bottom panels: evolution of hot loop as observed from IRIS 1330 \AA\ (Left) and AIA 94 \AA\ (right) filters. Both the time-distance images start from 03:09 UT.}
    \label{fig:hloop}
\end{figure*}

We further investigated the source of this small-scale brightening region with photospheric LOS magnetic field data. In the top panels of Fig.~\ref{fig:hbmevl}, we show images of evolution of magnetic field beneath the hot emission region. Brightening region is situated almost above the negative polarity region (blue in color) surrounded by the positive polarity regions (red in color). We obtained average positive and negative polarity magnetic field and flux evolutions underneath brightening region marked with a black box and plotted in left and middle panels of Fig.~\ref{fig:hbmevl}. Due to sensitivity limit on HMI LOS magnetic field data, we have chosen only pixels with a magnetic field strength above 10 G in to account. Total magnetic field strengths and fluxes decrease with time within the box region. Although box is dominated by negative polarity fields, there are elements of positive polarity fields also. A careful inspection of magnetic maps within the box area shows a strong region of magnetic field decay. This region is noted by the small yellow box within the bigger black box and associated magnetic field and flux evolution are plotted in the bottom right panel. In this box, average field strength (of negative polarity) decays from  210 G to 120 G within the span of about 40 min whereas that in bigger box decays from about 115 G to 100 G. Magnetic fluxes also evolve with time in the both boxes and show consistent decay of magnetic fluxes with a decay rate of about $10^{15}$ Mx s$^{-1}$. We thoroughly investigated flux decay in this region with different box sizes and found that observed decrease in fluxes are due to decay of magnetic fluxes and not due to any motion of magnetic patches out of the rectangular boxes considered here. We also noted that smaller box does not contain any positive polarity magnetic field pixel during the observed time interval. The decay of negative magnetic field strengths and fluxes can be interpreted as evidences of flux cancellations \citep[e.g.,][]{2018A&A...615L...9C,2019A&A...623A.176C}. 

As noted earlier, after the appearance of small-scale brightening region, loop structure was observed in the AIA passbands sensitive to hot plasma ( AIA 94, 131, and 335 \AA). Although loop became fully brightened up at different times in different filters with first appearance in hottest passband followed by cooler ones, similar to earlier reports \citep[e.g.,][]{2018ApJ...857..137G}. Additionally, here we also observed formation of loops at the both sides of transient brightening. In the top panel of Fig.~\ref{fig:hloop}, we show full extent of brightened loop as observed from IRIS-SJI 1330 \AA , and AIA 94 \AA\ passbands. We traced the hot loop from transient brightening region to another foot-point rooted in the neighboring sunspot and labeled them as points $A_H$ and $B_H$ respectively. On the other side of brightening region, many loops were formed which connected to different foot-points. These foot-points are located in the circular/elliptical shape with respect to each other and show some complex fine-threads connecting to each other. Due to the complexity of these loops and given the scope of current study, we will not be pursuing formation of this segment of loop here. We obtained the time-distance plot of hot loop from hot AIA 94 \AA\ and cool IRIS 1330 \AA\ passbands. Both the time-distance plots are presented in the bottom panels of Fig.~\ref{fig:hloop}. From the plots, it is clear that brightening front moves from foot-point A to B.

\begin{figure*}
	\includegraphics[width=0.9\textwidth]{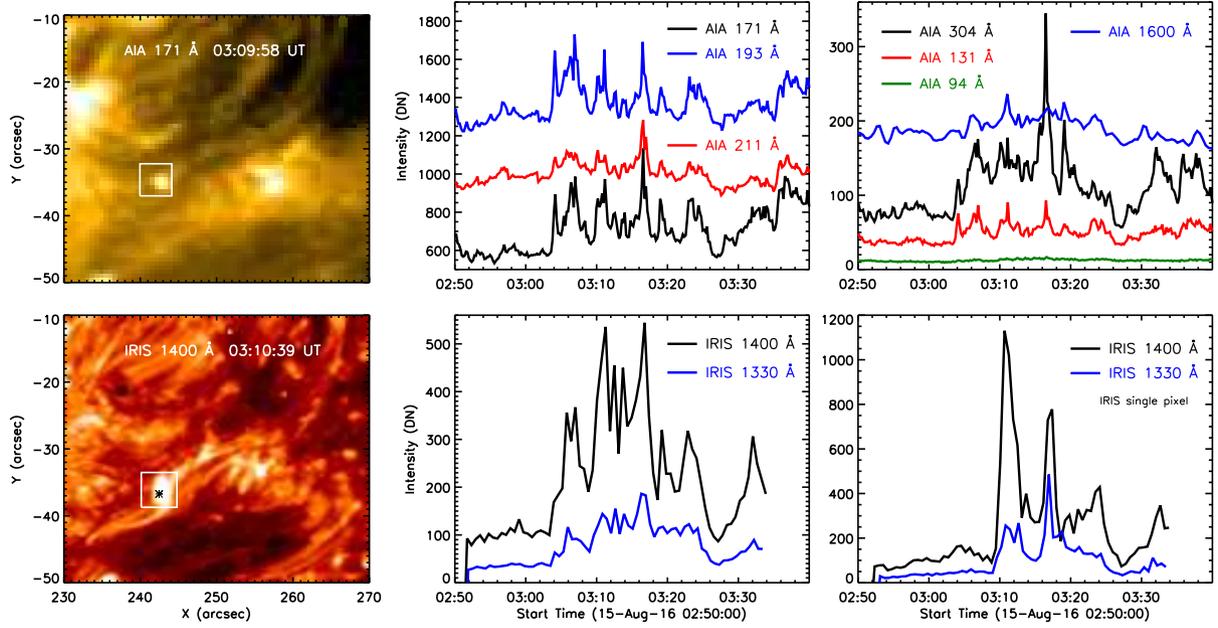}
    \caption{Different panels are same as in Fig.~\ref{fig:hbevl} but for cool loop transient.}
    \label{fig:cbevl}
\end{figure*}

Similar to hot loop, cool loop also appeared after the small-scale transient events. However, in this case we found several such transient events occurring at different locations during the course of whole cool loop formation. We studied three such events and found them similar to each other and chosen to present first event in detail. Location of this event is shown with a box on AIA and IRIS images in the left panels of Fig.~\ref{fig:cbevl} and time evolution of this region is plotted in right panels of Fig.~\ref{fig:cbevl} as labeled. Intensity evolution curves indicate that brightening in this region is intermittent in nature and has several peaks. In this case, we found strong variation in the intensity observed from AIA 304 \AA , 131 \AA , 171 \AA ,  193 \AA , and IRIS 1330 \AA\ and 1400 \AA\ filters whereas variation in other filters such as AIA 94 \AA , 335 \AA , 1600 \AA , and 1700 \AA\ were very weak. Such filter response suggest that this transient event released energy only up to the temperature $<2$ MK. Similar responses were also observed for associated loop which brightened up only in the same filters. Interestingly, no noticeable time delays were observed in any of these filters. In the bottom right panel of Fig.~\ref{fig:cbevl}, we also show IRIS light curves obtained at single pixel level. Location of this single pixel is  marked with asterisk (*) in the IRIS 1400 \AA\ image. These light curves show brightening enhancement of about $10-15$ times in the both IRIS 1330 and 1400 \AA\ filters. These findings suggest a clear distinction between the two type of transient events studied here. 

\begin{figure*}
	\includegraphics[width=0.75\textwidth]{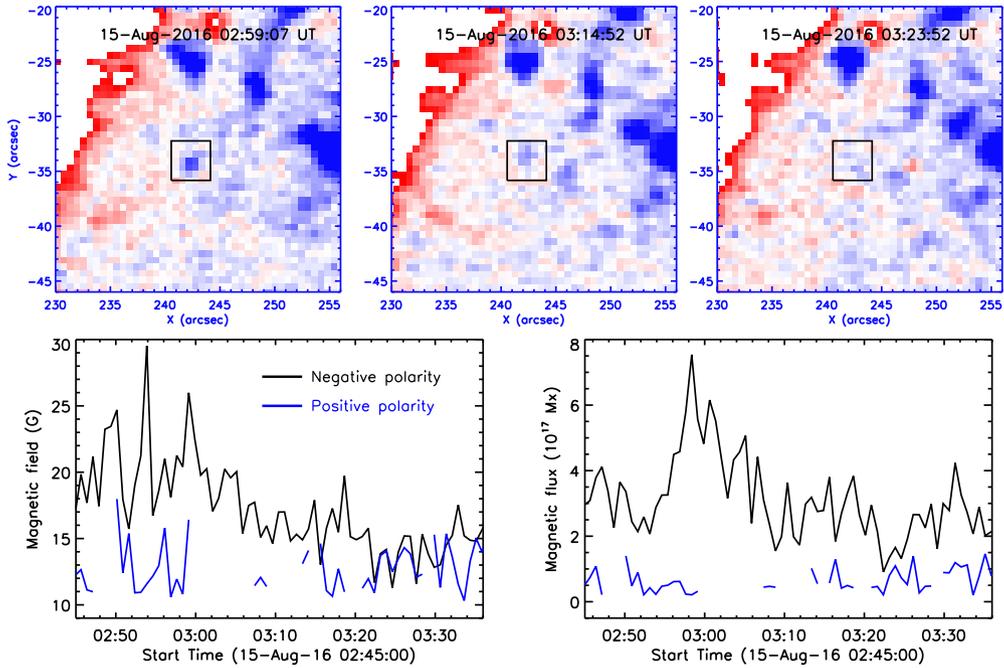}
    \caption{HMI LOS magnetic field images  ($\pm 200$ G) beneath cool loop transient. Description of different panels are same as in Fig.~\ref{fig:hbmevl}.}
    \label{fig:cbmevl}
\end{figure*}

We investigated the photospheric LOS magnetic field data under this cool transient which is a very weak mixed polarity region. In the top panels of Fig.~\ref{fig:cbmevl}, we plot images of evolution of magnetic field under this transient. White box represents location same as those chosen for brightening evolution in Fig.~\ref{fig:cbevl}. Magnetic field in this region is very weak, just above the sensitivity limit of HMI magnetogram of 10 G. We again obtained positive and negative polarity magnetic field and flux evolution in this box and plotted in the lower panels of Fig.~\ref{fig:cbmevl}. On an average, negative polarity fields and fluxes within the box decrease from $\approx 30$ to $ \approx 13$ G and 7.5$\times$10$^{17}$ to 1$\times$10$^{17}$ Mx respectively within the span of about 25 min. No inference can be drawn for the  positive polarity fields and fluxes due to weak strength and poor spatial resolution. However, observed flux cancellation of negative polarity field underneath this cool transient again provides evidence of magnetic reconnection to be the driver of such transients.

\begin{figure*}
	\includegraphics[width=0.98\textwidth]{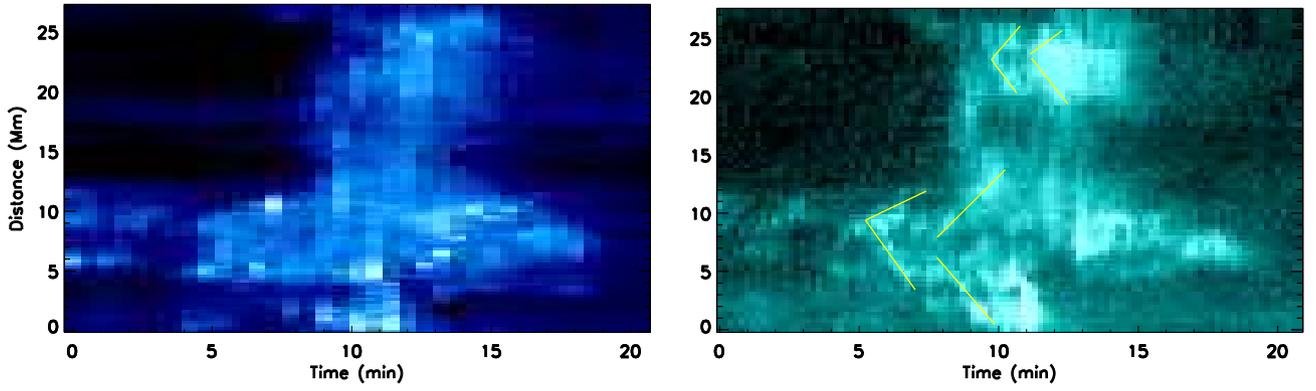}
    \caption{Evolution of cool loop observed from IRIS 1330 \AA\ (left panel) and AIA 131 \AA\ (right panel) filters from points A to B (see Fig.~\ref{fig:hloop}). Both the time-distance images start from 03:09 UT. Overplotted lines in the right panel indicate bidirectional flows from transient locations.}
    \label{fig:cloop}
\end{figure*}

\begin{figure*}
	\includegraphics[width=0.9\textwidth]{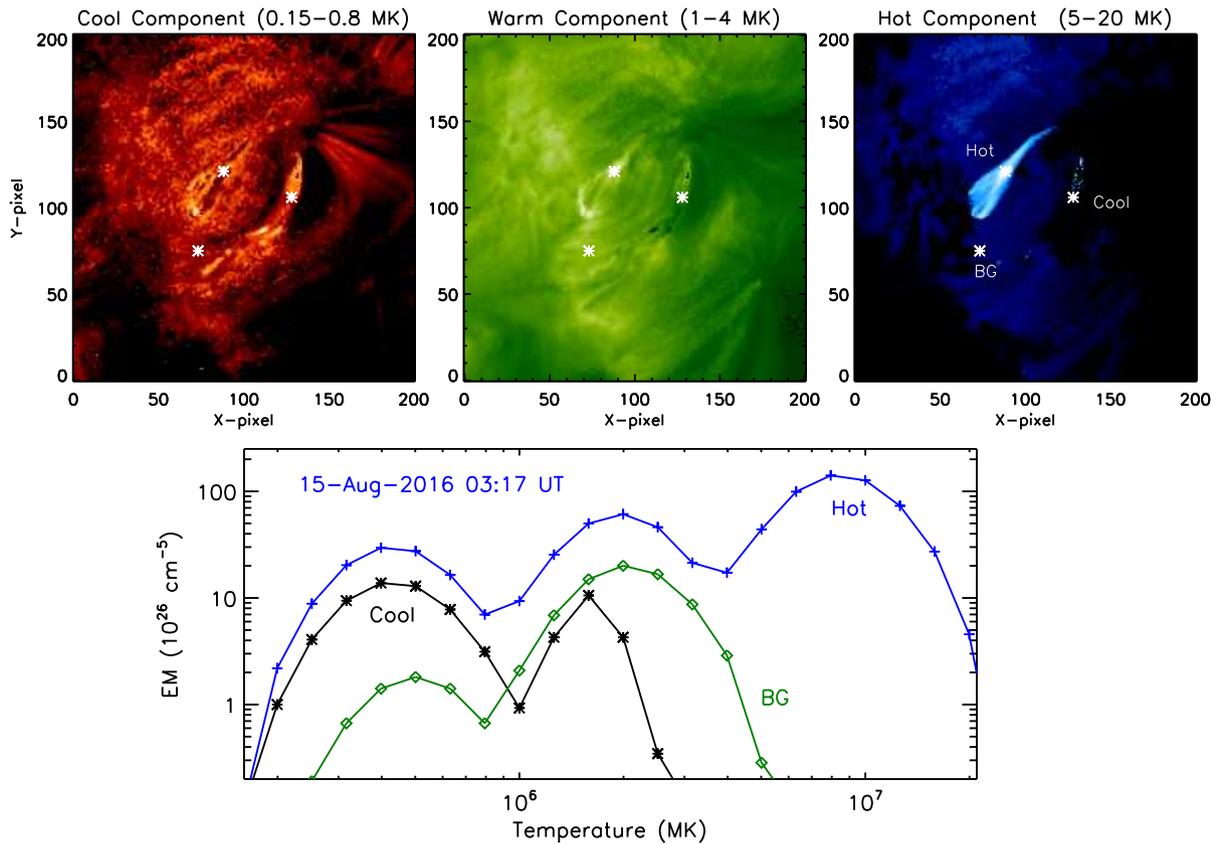}
    \caption{Top panels: Emission measure (EM) images obtained at 03:17 UT in different temperature bins as labeled. Lower panel: EM distribution curves for hot, cool, and background regions as labeled.}
    \label{fig:temp}
\end{figure*}

A fully brightened cool loop structure was visible in the IRIS filters (see Figs.~\ref{fig:context} and \ref{fig:hloop}). This loop was also visible in AIA 131 \AA\ passband but not in AIA 94 \AA . As AIA 131 \AA\ passband has strong contribution from Fe VIII 131 \AA\ line \citep{2010A&A...521A..21O}, appearance of this loop in AIA 131 \AA\ filter could be due to cool plasma components only.  We traced this cool loop from one extent ($A_C$) to another ($B_C$) reaching up to the sunspot boundary as shown in Fig.~\ref{fig:hloop}. This loop also showed some complex fine-threads within the loop segments, however, we again decided to treat them as single loop like-structure. Moreover, we also noticed that this loop has lot of mass motion towards the foot-point anchored in the sunspot region. We obtained time-distance plot of cool loop from  IRIS 1330 \AA\ and AIA 131 \AA\ filters and present it in Fig.~\ref{fig:cloop}. Plots clearly reveal several brightening fronts moving from both sides of transient events (marked with continuous lines). Several fine-threads within the loop structure were also noticed. These moving fronts and threads are resolved from high cadence data of AIA 131 \AA\ than from IRIS 1330 \AA\ images. 

\begin{figure*}
	\includegraphics[width=0.99\textwidth]{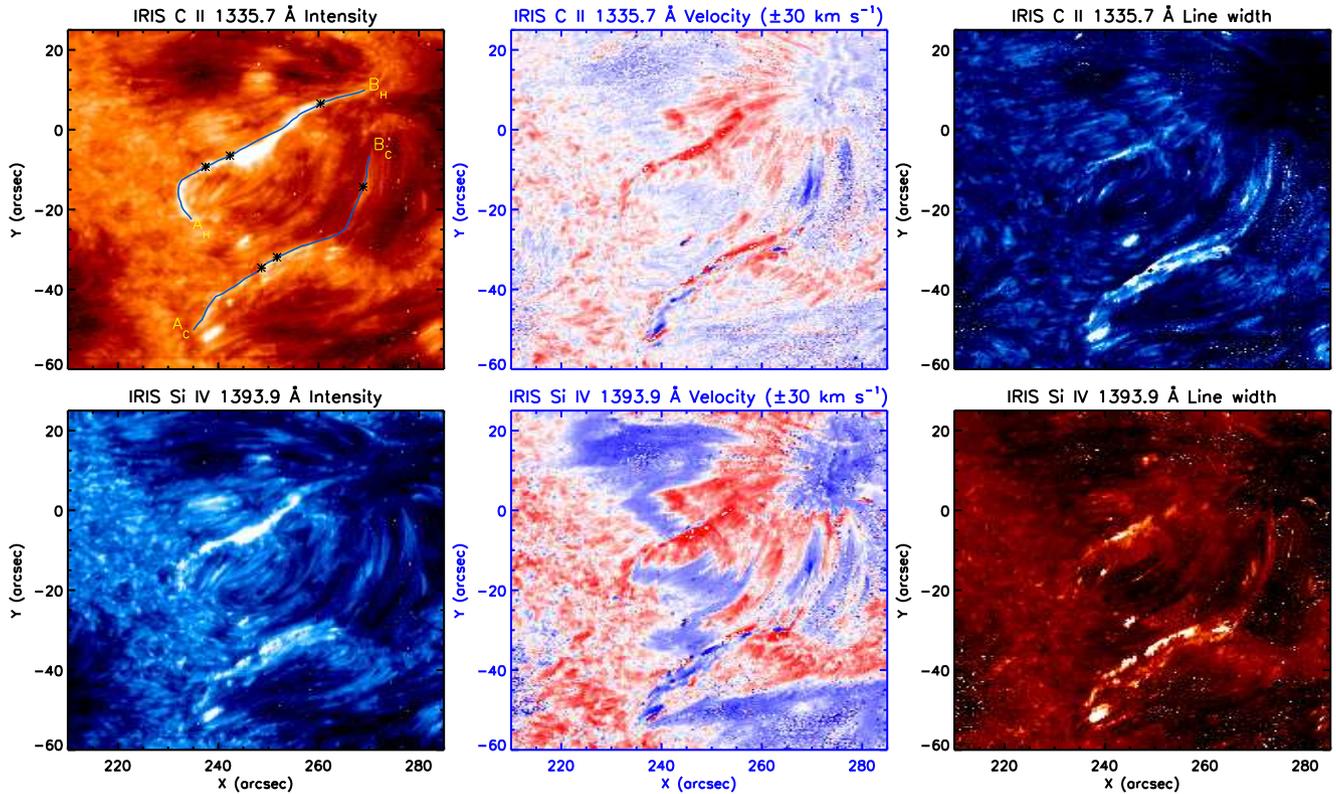}
    \caption{IRIS spectroscopic rastered images of hot and cool loops as obtained from intensity, Doppler velocity, and line width of C II 1335.7 \AA\ (top panels) and Si IV 1393.9 \AA\ (bottom panels) lines. Loops are traced from points A to B where subscripts H and C indicate hot and cool loops respectively.}
    \label{fig:spec_img}
\end{figure*}

To quantify the temperature of these loops, we employ the technique of Differential Emission Measure (DEM). For the purpose, we utilize DEM tool developed by \citet{2015ApJ...807..143C} using six coronal passbands of AIA 94 \AA , 131 \AA , 171 \AA ,  193 \AA , 211 \AA , and 335 \AA . This method provides estimate on amount of plasma present at different electron temperatures along the line-of-sight.

In Fig.~\ref{fig:temp}, we plot emission measure (EM) images at cool (0.15-0.8 MK), warm (1-4 MK), and hot (5-20 MK) plasma temperatures at time around 03:17 UT i.e. when both the loops were visible in AIA 131 \AA\ passband. Both hot and cool loop structures are clearly visible at cool plasma whereas only hot loop is visible at hot plasma. Nevertheless at warm plasma, both the loops are almost merged in to the background plasma except for few bright segments. Thus, EM images  indicate appearance of additional plasma components along these loops at cool and hot temperatures. For representation purpose, we plot EM curves at individual brightening locations of hot and cool loops in the bottom panel of Fig.~\ref{fig:temp}. We also plot nearby background region for comparison purpose. Locations are marked with asterisk (*) and labeled in EM image of hot plasma. EM curves clearly show appearance of additional hot (8 MK) and cool (0.4 MK) plasma  components in the hot loop whereas only cool (0.4 MK) plasma components in the cool loop as compared to background region. These distributions are almost consistent throughout the hot and cool loop lengths. Therefore, hot loop can be assumed to be composed of multi-thermal plasma around 8 and 0.4 MK while cool loop to be only around 0.4 MK.

We further obtained electron densities associated with these loops using EM values at peak temperatures and upon assuming plasma filling factor to be 1 (i.e. $n_e^{cor}\approx \sqrt{EM/l}$). For hot transient,  EM is about $140\times 10^{26}$ cm$^{-5}$ at 8 MK whereas for cool transient, EM is about $14\times 10^{26}$ cm$^{-5}$ at 0.4 MK. EUV brightenings associated with these transients were observed over the length scale ($l$) of about $3\arcsec$. Therefore, we estimated coronal densities of EUV brightenings of hot and cool transients to be about $8\times 10^{9}$ and $2.5\times 10^{9}$ cm$^{-3}$ respectively. These obtained densities are similar to coronal densities and thus are result of coronal response of the transients.

\subsection{Spectroscopic Analysis}
\label{sec:spectro}

Fortuitously, the spectroscopic slit of IRIS was rastering the region when both hot and cool loops were evolving. This allowed us to perform a detail spectroscopic study of both loops. We selected \ion{C}{II}~1335.7 {\AA} and \ion{Si}{IV}~1393.9 {\AA} spectra and fitted with single Gaussian functions to extract different line parameters for the representation purpose only. In Fig.~\ref{fig:spec_img}, we plot images of the region obtained in intensity, Doppler velocity, and line width from both lines. Intensity images clearly show that both the loops were well rastered by IRIS spectroscopic slit. Although lower part of cool loop is very well overlapped with rastering time while upper end started fading away during rastering. However, evolution of hot loops is very well captured by rastering slit. For our analysis purpose, we have traced both the loops again in rastered image from points A to B as shown in top left panel of Fig.~\ref{fig:spec_img}. This observation provides us an unique opportunity to compare various spectroscopic properties of both hot and cool loops all together. Although it should also be noted that different segments of both loops are scanned at different times, and thus represents different phases of loop evolution.

\begin{figure}
	\includegraphics[width=0.45\textwidth]{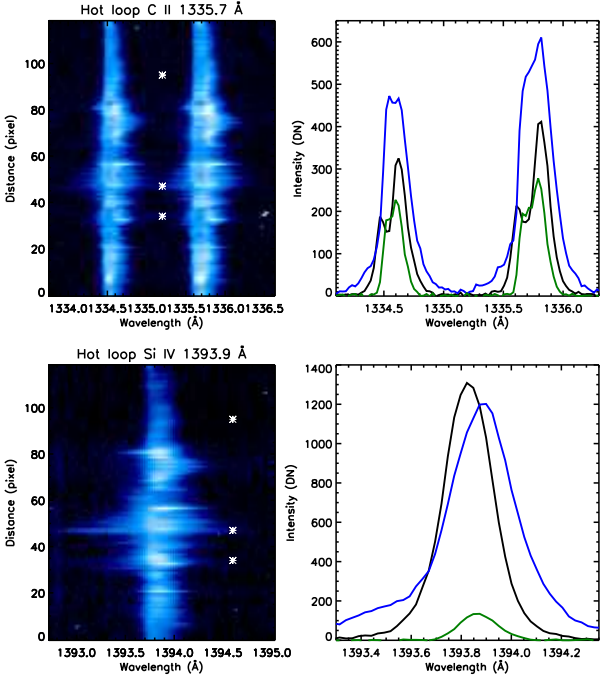}
    \caption{Left panels: wavelength-distance images obtained from \ion{C}{II}~1335 {\AA} (top) and \ion{Si}{IV}~1393.9 {\AA} (bottom) lines for hot loop traced from $A_H$ to $B_H$. Right panels: few examples of shape of spectral profiles along the hot loop. Locations are marked with asterisk (*) in the left panels.}
    \label{fig:hot-wvl}
\end{figure}

\begin{figure}
	\includegraphics[width=0.45\textwidth]{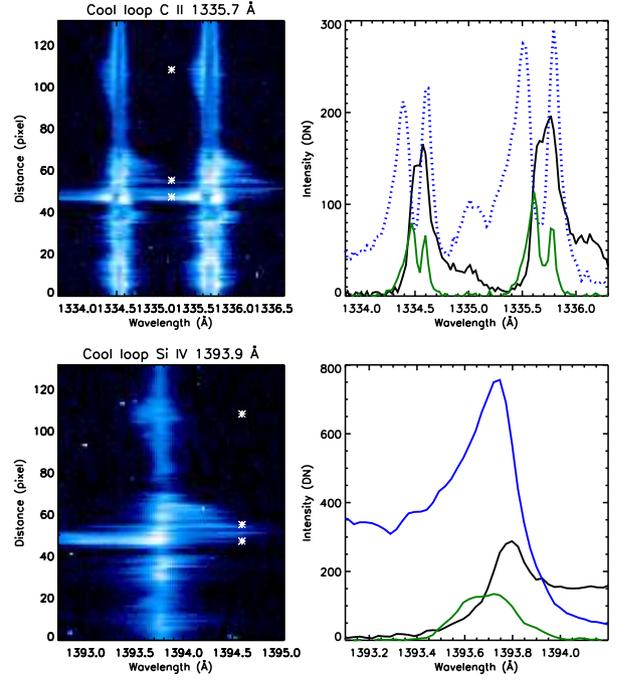}
    \caption{Same as in Fig.~\ref{fig:hot-wvl} but for cool loop.}
    \label{fig:cool-wvl}
\end{figure}

In Figs.~\ref{fig:hot-wvl} and \ref{fig:cool-wvl}, we show wavelength-distance plots (left panels) and few examples of shape of spectral line profiles (right panels) of hot and cool loops respectively. Plots clearly show several positions along the loops where spectral line profiles became highly broadened and took highly non-Gaussian shapes. For hot loop, excess emission in red and blue- wings of spectrum reached velocities more than $\pm 100$ km~s$^{-1}$. However wing enhancements reached beyond $\pm 200$ km~s$^{-1}$ for cool loop. Similarity between these profiles may suggest possibility of several such transients taking place along the length of these loops. These transients are connected to each other via formation of small-scale loops which when observed together appears as long loops, specially in the case of cool loop. Magnetic connectivity of these loops are studied in detail in Section~\ref{sec:nfff}.      

We have fitted double Gaussian function to line profiles shown in Figs.~\ref{fig:hot-wvl} and \ref{fig:cool-wvl}. Primary Gaussian represents steady emission with contribution from background and foreground. Whereas secondary broad Gaussian function was included to account for additional component arising due to transient brightenings and loop formation. 

\begin{figure}
	\includegraphics[width=0.48\textwidth]{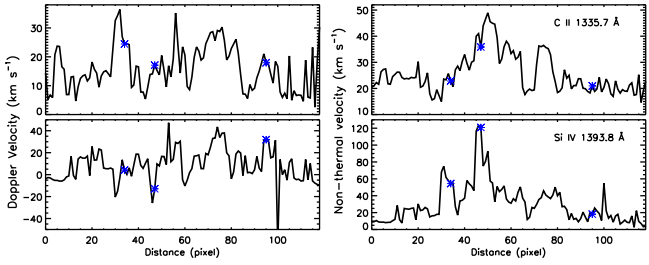}
    \caption{Variation of Doppler velocity (left panels) and non-thermal velocity (right panels) obtained from C II 1335.7 \AA\ and Si IV 1393.8 \AA\ lines along the hot loop (from point A to B). Location of asterisks (*) are same as those in Fig.~\ref{fig:hot-wvl}.}
    \label{fig:hparam}
\end{figure}

\begin{figure}
	\includegraphics[width=0.48\textwidth]{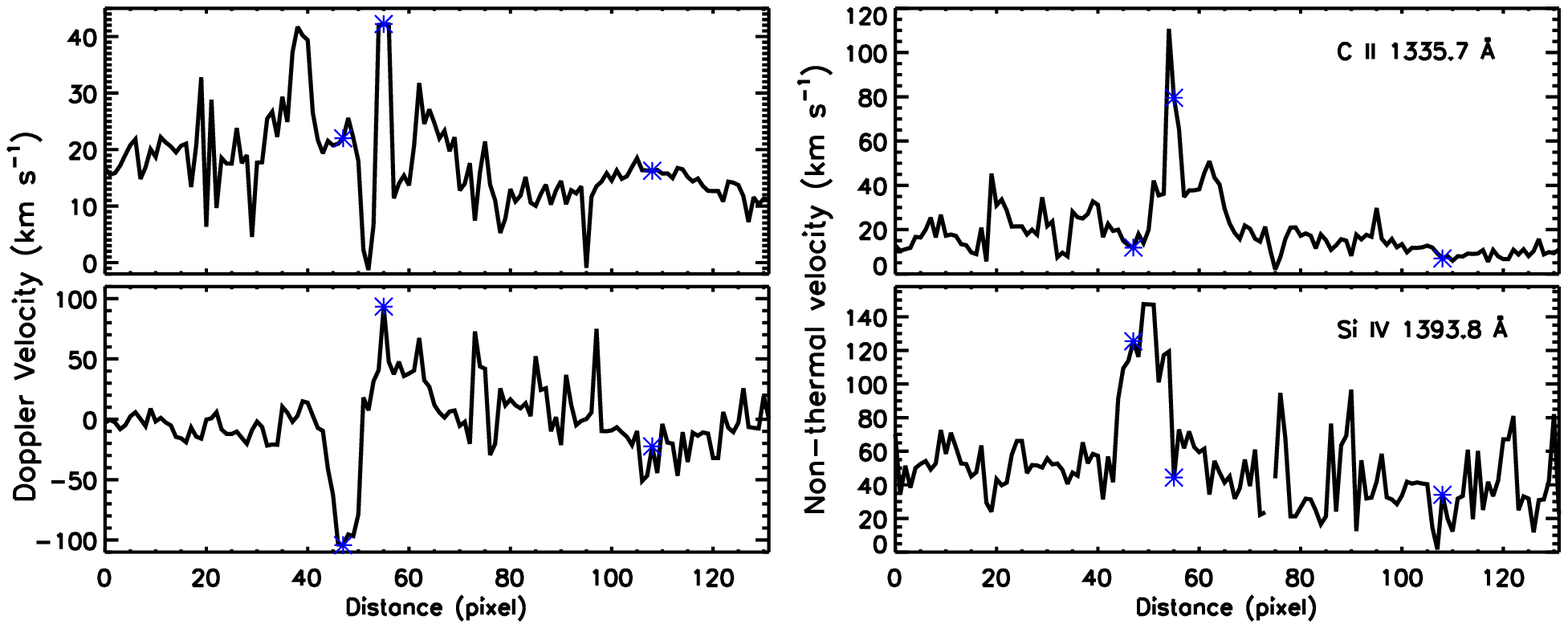}
    \caption{Panels are same as in Fig.~\ref{fig:hparam} but for cool loop.}
    \label{fig:cparam}
\end{figure}

In Figs.~\ref{fig:hparam} and \ref{fig:cparam}, we plot Doppler velocity (left panels) and non-thermal velocity (right panels) along the hot and cool loops as obtained from secondary Gaussian function of \ion{C}{II} 1335.7 and \ion{Si}{IV} 1393.8 \AA\ lines. During the hot loop evolution, both \ion{C}{II} 1335.7 and \ion{Si}{IV} 1393.8 \AA\ lines show predominantly red-shifts i.e. down-flow of more than 10 km~s$^{-1}$. There are some locations where down-flow velocities can reach more than 20 km~s$^{-1}$ as obtained from \ion{C}{II}. Whereas such enhancements in velocities of 20 km~s$^{-1}$ were found in both downward (red-shift) and upward (blue-shift) directions from \ion{Si}{IV}. We also noticed that velocities obtained from \ion{Si}{IV} is predominantly larger than that obtained from \ion{C}{II}. Moreover, cool loop also shows predominantly red-shifts i.e. down-flow of about 10-15 km~s$^{-1}$ from \ion{C}{II}. Whereas \ion{Si}{IV} shows mixed pattern of up- and down-flows with velocities reaching more than $\pm 80$ km~s$^{-1}$ at some locations. These locations correspond to transient events which occurred during the course of cool loop evolution.

In the right panels of Figs.~\ref{fig:hparam} and \ref{fig:cparam}, we also plot non-thermal velocities along the both loops. Average non-thermal velocities estimated from \ion{Si}{IV} line is greater than those estimated from \ion{C}{II} line for both the loops. However, non-thermal velocities obtained from \ion{Si}{IV} line along cool loop is predominantly larger than that obtained along hot loop. Whereas non-thermal velocities measured from \ion{C}{II} along the hot loop is larger than that along the cool loop.  Moreover, average non-thermal velocities obtained from \ion{C}{II} and \ion{Si}{IV} lines at the location of transients along cool loop reaches more than 40 (maximum 110) km~s$^{-1}$ and 90 (maximum 148) km~s$^{-1}$  respectively. Whereas for hot loop those reaches only about 35 (maximum 49) km~s$^{-1}$ and 65 (maximum 123) km~s$^{-1}$ for the both lines respectively. 

During the raster scan, \ion{O}{IV} 1399.78 and 1401.16~{\AA} (peak formation temperature $\approx 0.14$ MK) lines also appeared in the spectra of both loops. Intensity ratio of this line pair is sensitive to electron density and thus provided us an opportunity to estimate electron densities along the both loop lengths. As both the lines are not strong enough to fit the Gaussian function, we integrated the intensity counts of both line profiles within -40 to 64 km~s$^{-1}$ with respect to their theoretical line centers and obtained the total intensity counts for both lines. This asymmetry in profile summation is primarily due to red-shifted nature of spectral lines (see left panels of Figs.~\ref{fig:hparam} and \ref{fig:cparam}). Although \ion{O}{IV} lines are found to be blended with other cooler lines \citep{2015arXiv150905011Y}, however in this observation, we did not find any noticeable blending upon inspecting detector image around both the lines which were visible in the Fig.~2 of \citet{2015arXiv150905011Y}.

\begin{figure}
	\includegraphics[width=0.5\textwidth]{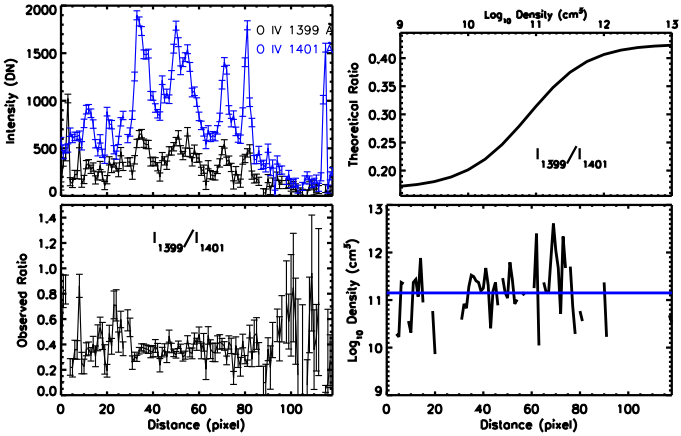}
    \caption{Left panels: variation of intensity (top panel) and intensity ratio (bottom panel) obtained from O IV 1399.78 and 1401.16 \AA\ lines along the hot loop (from point A to B). Right panels: theoretically predicted intensity ratio with electron number density (top panel) and estimated number density based on observed ratio along the hot loop (bottom panel). Average number density along the loop is also over-plotted with continuous blue line.}
    \label{fig:hdens}
\end{figure}

\begin{figure}
	\includegraphics[width=0.5\textwidth]{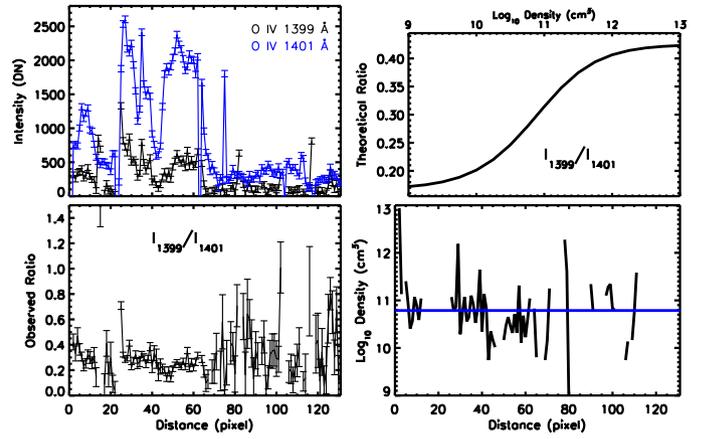}
    \caption{Panels are same as in Fig.~\ref{fig:hdens} but for cool loop.}
    \label{fig:cdens}
\end{figure}

 In Figs.~\ref{fig:hdens} and \ref{fig:cdens}, we plot variation of intensities obtained from \ion{O}{IV} 1399.78 and 1401.16~{\AA} lines along the loop lengths (top left panels). We also plot observed ratio of \ion{O}{IV} $\lambda1399/\lambda1401$ line pair along the loop length in bottom left panels. In top right panels, we plot theoretically predicted line ratios of \ion{O}{IV} $\lambda1399/\lambda1401$ with respect to electron densities as obtained from CHIANTI solar soft distribution \citep{1997A&AS..125..149D,2021ApJ...909...38D}. Using these ratios and densities, we deduced electron densities along the loops from observed intensity ratios. Estimated electron densities along hot and cool loops are plotted in the bottom right panels of Figs.~\ref{fig:hdens} and \ref{fig:cdens} respectively. At several locations, we found that observed ratios were going beyond the limit provided by theoretically predicted ratio of \ion{O}{IV} $\lambda1399/\lambda1401$ line pair, and thus did not plot data-points at those locations. However for our study purpose, we obtained average electron densities along the hot and cool loops and found them to be around $10^{11.2}$ and $10^{10.8}$ cm$^{-3}$ respectively. These density estimates suggest that both hot and cool loops are situated at very low down in the solar atmosphere, most likely at the upper chromospheric heights. Although it should also be noted that these theoretical ratios of intensities with electron densities are obtained under the assumption of optically thin plasma and established thermal equilibrium conditions. However, these events are transient in nature and are occurring at much lower in the solar atmosphere where these conditions may not be fulfilled. Therefore, these electron density estimates should be taken with caution.

\subsection{Energetics of transients}
\label{sec:energy}

Transient loops under study have distributions of electron densities and temperatures as obtained from line ratio and DEM techniques. In Table~\ref{tab:summary}, we provide summary of all the parameters obtained for both the loops. Henceforth, we calculate thermal energy $E_t\approx 3 n_e k_B T_e l^3$ (where l$\approx 3\arcsec$ is length scale of EUV brightening) at different temperatures at the location of transients. Upon using all the estimates, we find that thermal energies released at EUV wavelength ($T_e \approx 8$ MK) is $ \approx 3\times 10^{26}$ erg and at UV wavelength is $ \approx 2\times 10^{26}$ erg for hot transient. For cool transient, released thermal energies are $ \approx 4\times 10^{24}$ erg and $ 6\times 10^{25}$ erg at EUV ($T_e \approx 0.4$ MK) and UV wavelengths respectively.

We also obtained turbulent energy $E_{turb}\approx 3 n_e m_p v_{turb}^2  l^3$ at transition region temperature ($T_e \approx 80$ kK) using average non-thermal velocities obtained from Si IV 1393.7 \AA\ line. We identified locations of transients wherever there were significant enhancements in the non-thermal velocities along both loops. This provides an average non-thermal velocity to be about 60 and 100 km~s$^{-1}$ for hot and cool transients respectively. From these, we find turbulent energies at transition region temperature to be  about $3.7\times 10^{26}$ erg and $3.2\times 10^{26}$ erg respectively. Thus, the total observed energies released  were of the order of $9\times 10^{26}$ erg and $4\times 10^{26}$ erg at hot and cool transients respectively.

We also estimate amount of magnetic energy available at the sites of hot and cool transients. We find that magnetic flux under the hot transient decays from $17\times 10^{19}$ Mx to $15.1\times 10^{19}$ Mx within the bigger box of Fig.~\ref{fig:hbmevl}. Whereas it decays from $7.5\times 10^{17}$ Mx to $1\times 10^{17}$ Mx under the cool transient as seen in Fig.~\ref{fig:cbmevl}. Therefore, canceled flux ($\phi$) is estimated to be about $1.9\times 10^{19}$ Mx and $6.5\times 10^{17}$ Mx for hot and cool transients respectively. We thus estimate amount of magnetic energy converted during the flux cancellation by $E_{mag}=\frac{1}{8\pi} \phi^2 h/a$ where $a$ is area of EUV brightening ($\approx 3\arcsec \times 3\arcsec$ for both events) and $h$ ($\approx 500$ km) is chromospheric heating scale height as suggested by \citet{2018A&A...615L...9C} and references therein. This provides available magnetic energy to be about $1.5\times 10^{28}$ erg and $1.7\times 10^{25}$ erg at hot and cool transients respectively. Comparing these with the total energies released indicate that magnetic energy is sufficient enough to drive hot transient, however additional energy is needed to drive cool transient. This additional magnetic energy seems to be invisible here due to poor sensitivity and spatial resolution of HMI magnetogram.

\begin{table}
	\centering
	\caption{Summary of parameters obtained for hot and cool loops as observed from AIA and IRIS}
	\label{tab:summary}
	\begin{tabular}{lcc} 
		\hline
		 & Hot loop & Cool loop \\
		\hline
AIA life time (min)                   &  10--11                  &   6--7                  \\
AIA-DEM hot $T_e^h$ (MK)              &   8                      &    0.4                   \\
AIA-DEM $N_e^h$  (cm$^{-3}$)          &   8$\times$ 10$^9$       & 2.5$\times$ 10$^9$        \\
IRIS O IV cool $T_e^c$  (MK)          &   0.14                   &   0.14                     \\
IRIS O IV $N_e^c$ (cm$^{-3}$)         &   1.6$\times$ 10$^{11}$  &  6.3$\times$ 10$^{10}$      \\
Dopp. vel. $v_d$ (km s$^{-1}$) C II   &  10                      &  10-15                  \\
\hspace*{3cm}                 Si IV   &  10-15                   &  $\pm$ 10-15            \\  
Turb. vel. $v_{turb}$ (km s$^{-1}$) C II  &  35   & 40  \\
\hspace*{3.3cm}                 Si IV   &    60   &  100      \\
		\hline
	\end{tabular}
\end{table}

\begin{table}
	\centering
		\caption{Summary of different energies involved in the hot and cool transients}  
	\label{tab:energy}
	\begin{tabular}{lcc} 
		\hline
		 & Hot transient & Cool transient \\
		\hline
Thermal energy at EUV (erg)     &    $3\times 10^{26}$    &   $4\times 10^{24}$      \\
\hspace*{2.10cm} at UV ~~(erg)  &    $2\times 10^{26}$    &   $6\times 10^{25}$      \\
Turbulent energy at UV (erg)    &  $3.7\times 10^{26}$    &  $3.2\times 10^{26}$     \\
Observed total energy ~~~(erg)  &    $9\times 10^{26}$    &   $4\times 10^{26}$       \\
Available magnetic energy (erg)           &  $1.5\times 10^{28}$    & $1.7\times 10^{25}$      \\
		\hline
	\end{tabular}
\end{table}

\section{The NFFF extrapolation model}
\label{sec:nfff}

\begin{figure*}
\includegraphics[width=0.4\linewidth]{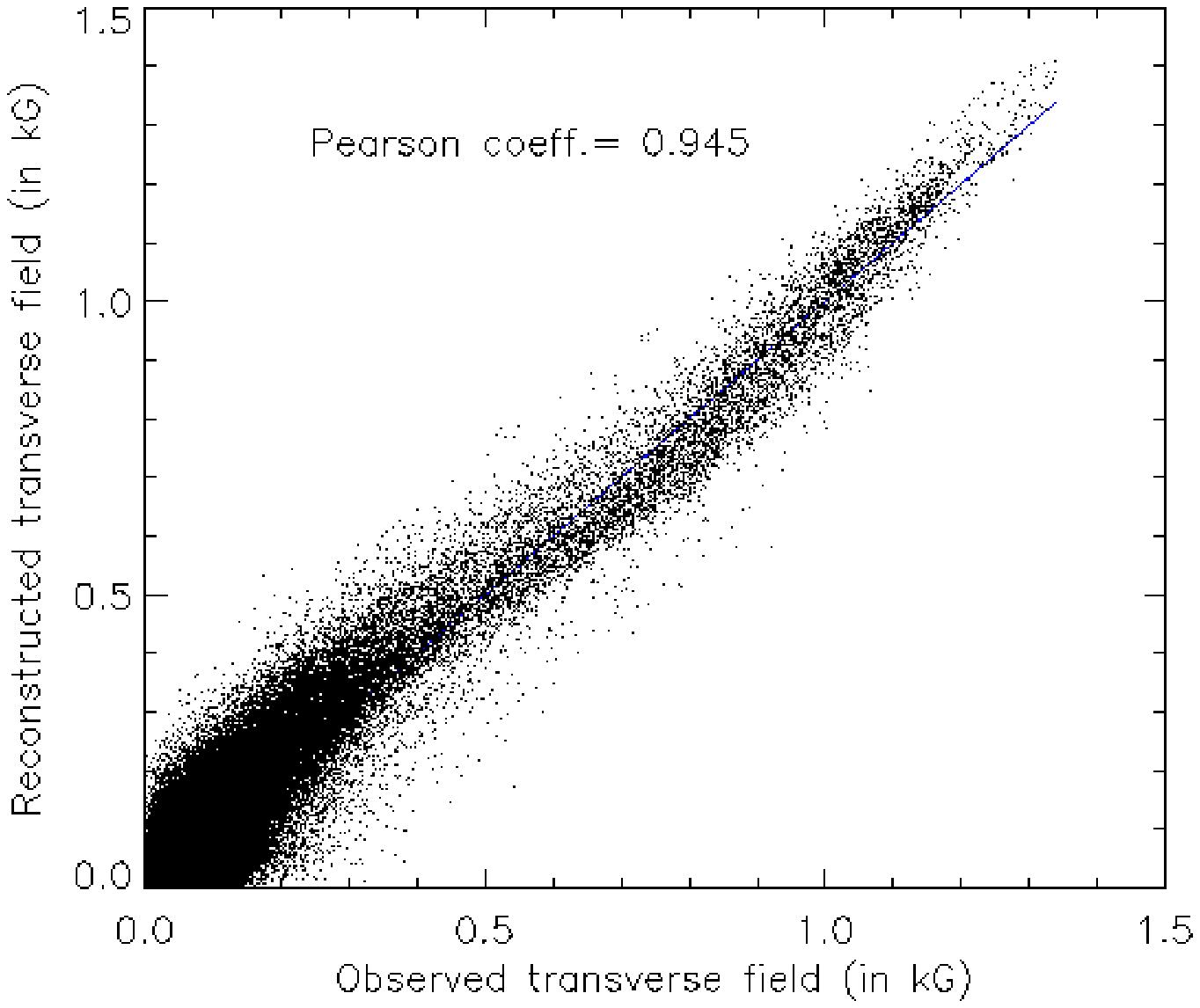}\includegraphics[width=0.42\linewidth]{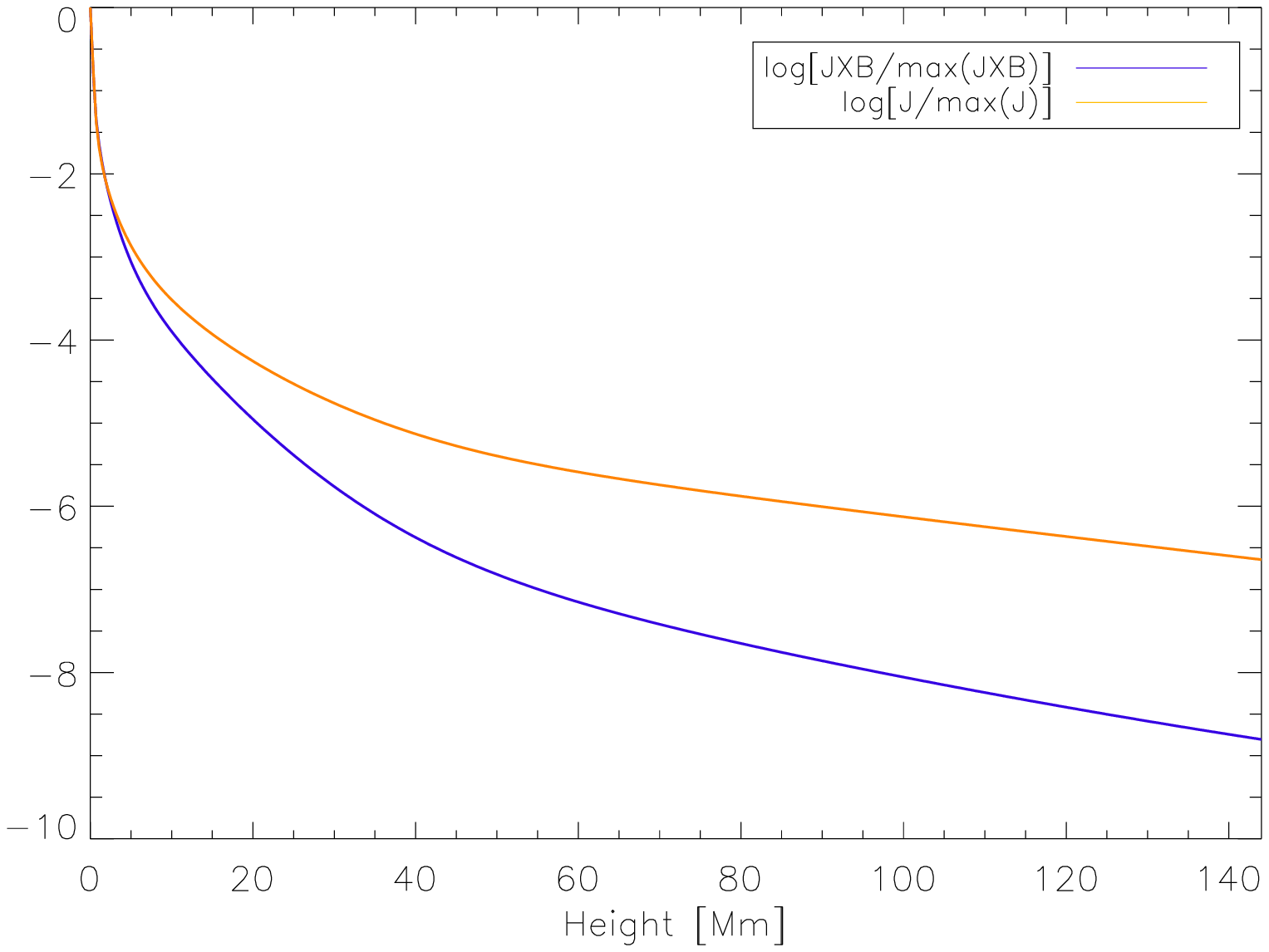}
\caption{Left panel: Pearson correlation coefficient obtained between observed and reconstructed transverse components of magnetic field on the photosphere. Right panel: Variations of Lorentz force and current density with height.}
\label{fig:current}
\end{figure*}

\begin{figure*}
\includegraphics[width=0.75\textwidth]{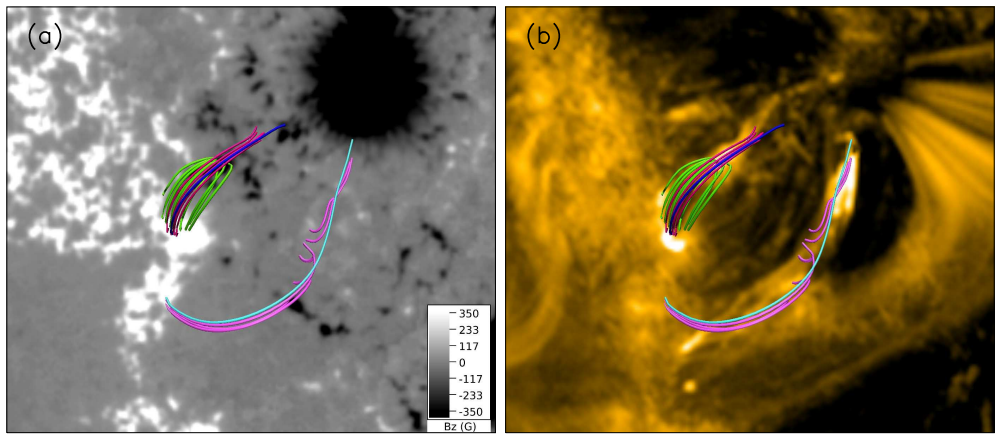}
\caption{Panel (a): Different extrapolated magnetic field topology near the hot and cool loops plotted over HMI LOS magnetogram data ($\pm 350$ G). Panel (b): Extrapolated field lines are plotted over AIA 171 {\AA} image.}
\label{fig:extrapolated}
\end{figure*}

To understand magnetic field topology of transient regions and associated hot and cool loops, we performed magnetic field extrapolation using available photospheric magnetogram from HMI. For the purpose, we utilized  Non-force-free-field (NFFF) extrapolation model. It assumes the photosphere non-force-free \citep[since only the mid-corona is force-free,][]{2001SoPh..203...71G} while the Lorentz force decays with height. The details of NFFF model can be found in \citet{2007SoPh..240...63B} and \citet{2008SoPh..247...87H}. To name a few, the model has constructed the coronal magnetic field topology successfully in earlier studies like the formation of a circular flare ribbon \citep{2018ApJ...860...96P} and an eruption of a blow-out jet \citep{2019ApJ...875...10N}.

We utilized \textquoteleft hmi.sharp.cea.720s\textquoteright\ series magnetogram obtained from HMI/SDO at 03:12 UT ($\approx 5$-minutes prior to the peak of transient brightenings). Magnetogram is processed by the Lambert equal area projection and is flux balanced \citep{1990SoPh..126...21G}. Spatial extensions of the magnetogram are $790\times 560$ pixels along $x$ and $y$ axes in a Cartesian coordinate system. Corresponding physical extensions are $\approx 284$ Mm in $x$-axis, $\approx 201$ Mm in $y$-axis, and $\approx 144$ Mm in $z$-axis. The minimized error in the extrapolation, denoted as $E_{n}$ is saturated to $\approx 0.262$, after 2000 number of iterations. Further, we have calculated Pearson correlation factor between the two transverse components on the photosphere and found a strong value of 0.945 (left panel of Fig.~\ref{fig:current}). To appreciate the usage of NFFF, we have shown logarithmic change of Lorentz force and current on each layer (normalized to their maximum value) in the computational box with height in the right panel of Fig.~\ref{fig:current}. Notable is the fast decrease of Lorentz force than current that effectively makes corona almost force free, adhering to the usual trend seen in the solar atmosphere, and thus validates our choice of extrapolation model.  

In Fig.~\ref{fig:extrapolated}, we plot magnetic field topologies in the vicinity of both hot and cool loops as identified in Figs.~\ref{fig:context} and \ref{fig:hloop}. In panel-(a), extrapolated field lines depicted in separate colors show different connectivities in the regions and are easily identifiable with respect to background image of B$_z$ component of the magnetic field. Similarity between the extrapolated field lines and the loops observed in AIA 171 \AA\ passband can be seen in the panel-(b). Near the hot loop region, the green color field lines depict connectivities from small transient region (marked by white box in Fig.~\ref{fig:hbevl}) to extended positive polarity. This is the positive polarity where several connectivities were observed in the circular/elliptical shape in AIA 94 \AA\ image (see Fig.~\ref{fig:context}).  Single blue color field line shows connectivity from small negative polarity slightly away from transient region in the top right side which is situated along the hot loop to again on the extended positive polarity on left side. No small magnetic field lines were found to connect hot loop to sunspot region as there are hardly any positive polarity field present in the vicinity of sunspot region (which is negative in polarity) as observed with the spatial resolution of HMI magnetogram. Near cool loop region, pink color field lines denote inter-connectivities from negative polarity to extended positive polarity in steps of multiple small loops. These small loops are connecting several opposite polarity magnetic patches and are following the observed cool loop except near the lower end of cool loop. Near the lower end of cool loop, no positive polarity patches were observed with HMI magnetogram which can lead to any magnetic field connectivities as those observed in AIA 171 \AA\ image (panel-(b) of Fig.~\ref{fig:extrapolated}). Cyan color field line shows a long connection between the two main polarities, although no such loop is observed in AIA 171 \AA\ image.

\section{Discussion and Summary}
\label{sec:discussion}

In this paper, we presented a detailed spectroscopic and imaging study of two nearby transient loops formed around a sunspot region using IRIS and AIA-HMI/SDO observations. Loops were termed as hot and cool loops based on their temperature distributions. Loops were associated with small-scale transients which occurred before the loops were brightened up. Evidence of magnetic flux cancellation were found underneath these transients. Loops and transients are multithermal in nature and have chromospheric to coronal density and temperature distributions as summarized in Table~\ref{tab:summary}.

\begin{figure}
\centering
	\includegraphics[width=0.40\textwidth]{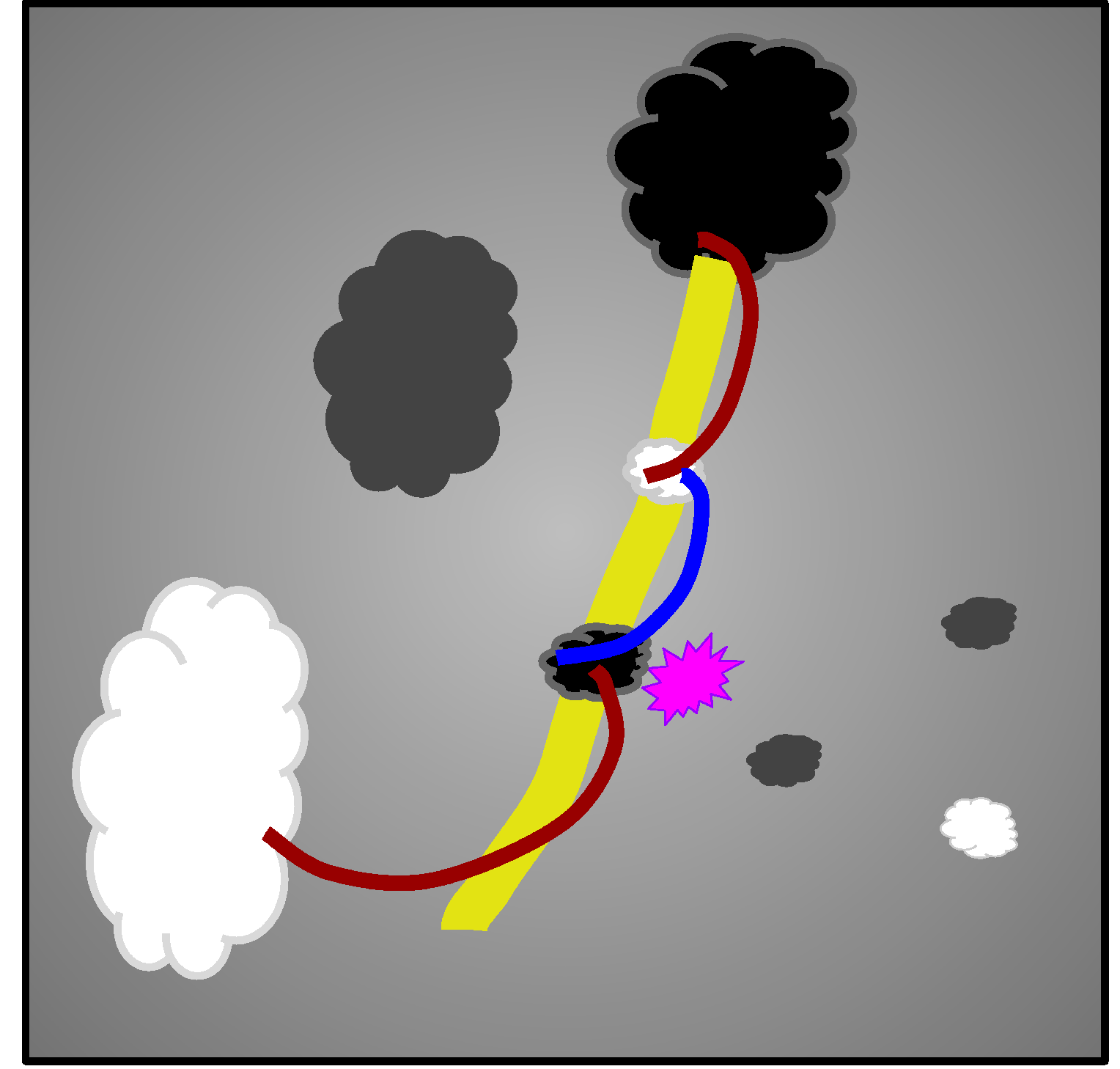}
    \caption{Cartoon depicting magnetic structure of cool loop and associated small-scale transient. Transient is a result of component wise reconnection taking place between field lines emanating from same polarity region.}.
    \label{fig:cartoon}
\end{figure}

Observations of magnetic flux cancellations on the photosphere can be considered as a signature of magnetic reconnections in the solar atmosphere which powers such transients \citep{2018ApJ...862L..24P,2018MNRAS.479.2382L}. Although on a very small spatial scale, we didn't find any presence and decay of opposite polarity fluxes but it is likely that magnetic fluxes at opposite-polarity cancellation sights are much more dispersed in nature and below the detection limit of HMI instrument. This has been clearly showed by \citet{2017ApJS..229....4C} using LOS magnetogram from IMaX which has six times better spatial resolution than HMI. Their observations revealed several small-scale opposite-polarity magnetic elements close to the loop foot-points of dominant polarity which were not visible in HMI magnetogram. Therefore, in our observations also it is highly plausible that flux cancellation of positive polarity magnetic field elements on a small scale is hidden from the HMI spatial resolution. Moreover, magnetic loops can also interact by component reconnection mechanism at higher heights. Recently similar component reconnection is also proposed for small-scale brightening events, particularly for the campfires observed by the Solar Orbiter \citep{2021A&A...656L...7C}. For our study, we have illustrated a plausible situation  in Fig.~\ref{fig:cartoon} where the magnetic field lines can interact with each other and heat the plasma via component reconnection in the atmosphere. Though the configuration matches with the cool loop, the same mechanism may also act in case of hot loop. However in such configurations, transient events will mainly be restricted to the region of closest approach between the interacting field lines which will be in this case near the magnetic patch itself. This will also explain almost co-spatial location of transient events and region of flux cancellation of one of the magnetic polarity. After reconnection, hot plasma moves in the opposite direction as seen in Fig.~\ref{fig:cloop}. Moreover, it should also be noted that simulation results of \citet{2014Sci...346B.315T} showed the interaction of non-thermal electrons accelerated in small heating events with lower solar atmosphere can also reproduce brightenings observed in the IRIS data similar to our observations. However, our observations of decay of magnetic fluxes at the co-spatial transient locations favor magnetic reconnection to be the primary cause of observed brightenings as also suggested by \citet{2018A&A...615L...9C}.

For cool loop, several transient brightenings were observed. Loop is composed of several small-scale fine threads which were observed from IRIS-SJI 1400 \AA\ images similar to \citet{2015ApJ...810...46H} and also from magnetic field extrapolation plots (see Fig. \ref{fig:extrapolated}). Electron densities obtained from IRIS transition region lines were more than a magnitude larger than those obtained from AIA DEM analysis using coronal lines. This clearly suggests that loop threads observed at different temperatures are not spatially identical at coronal and transition region temperatures. Recently, \citet{2021ApJ...909..105T} also suggested multi-stranded or different strand structures for hot transient loops formed in the active region core based on their observations of AIA 304 and 94 \AA\ images. Henceforth, our findings also suggest multi-stranded nature for not only hot loop and transient but also for cool loop and transient.

 Densities obtained here for cool loop is quite higher than those reported by \citet{2015ApJ...810...46H} of about $10^{10}$ cm$^{-3}$ at transition region temperature. However, densities are comparable with findings of \citet{2018ApJ...869..175H} for a cool loop reaching temperature up to 2 MK. For hot loop, \citet{2018ApJ...857..137G} found electron densities to be  $10^{11.5}$ cm$^{-3}$ for transition region temperature whereas \citet{2020ApJ...902...31T} found densities reaching of the order of $10^{10}$ cm$^{-3}$ at hot coronal temperatures. These reported densities for hot loops are similar to our findings. All these reported densities suggest that these transients are occurring lower in the solar atmosphere, most likely in the chromosphere. However our findings also suggest that these hot loop and associated transients are formed slightly lower down in the solar atmosphere as compared to cool loop and associated transient based on their both transition region and coronal density estimates.

Spectroscopic observations also provided Doppler and non-thermal velocities along both hot and cool loops and are also summarized in Table~\ref{tab:summary}. Both \ion{C}{II} and \ion{Si}{IV} lines showed predominantly down-flow of about 10 km~s$^{-1}$ along the hot loop. Observations of such down-flows of lower atmospheric lines are very common at flare foot-points \citep[e.g.,][]{2006SoPh..239..173D,2006ApJ...638L.117M} and also in micro-flares \citep[e.g.,][]{2018ApJ...857..137G}. Flow patterns along cool loop were quiet complex and showed mix of both up- and down-flows as observed from \ion{Si}{IV} line. Close inspection of middle panels of Fig.~\ref{fig:spec_img} shows fine-threads in cool loop showing evidence of anti-parallel flows at some locations with magnitude larger than 25 km~s$^{-1}$ and 50 km~s$^{-1}$ for \ion{C}{II} and \ion{Si}{IV} lines respectively. Such oppositely directed flows were also observed in slit-jaw images, however due to poor cadence of observations, their quantification was not feasible. Moreover, these flows appeared to originate from transients occurring along the cool loop. \citet{2013ApJ...775L..32A} has also found several such anti-parallel flows along adjacent AR filament threads observed with High-resolution Coronal Imager (Hi-C). Temperature of such threads were found to be 0.28 MK from AIA DEM analysis whereas in our observation, cool loop has temperature of about 0.4 MK. These results suggest that such anti-parallel streaming flows are quite common within fine filament-threads \citep[e.g.][]{2013ApJ...775L..32A}. However, \citet{2015ApJ...810...46H} noted that such anti-parallel flows can also be a result of projection effect and 3D geometry of the loops. Therefore, a detailed investigation is required on the anti-parallel flows along cool loop fine-threads. 

Spectral profiles of \ion{C}{II} and \ion{Si}{IV} lines along hot and cool loops showed highly non-Gaussian and broad shape at several locations (see Figs.~\ref{fig:hot-wvl} and \ref{fig:cool-wvl}). These unresolved motions though require lot of energy and momentum to reach such high velocities. Similar profiles at several locations along the cool loop can easily result in oppositely directed plasma flows as discussed above. Several such events in the close vicinity can cause oppositely directed flows along the fine threads as observed in the Doppler velocity images (see middle panel of Fig.~\ref{fig:spec_img}). Possibility of some fine-threads to be very close by without being resolved by IRIS raster can also explain some of the highly broadened line profiles along the loops. Few of such examples were also found by \citet{2015ApJ...810...46H} in their cool loops study. Recent 2-D numerical simulations of \citet{2020ApJ...894..155S} suggested that formation of cool loop in chromosphere and transition regions is a result of velocity response of transient energy released. These transients are due to explosive events or IRIS bursts above the cool loop footpoints. Our findings are in accordance with their simulation results.

All the energies involved during the transients are summarized in Table~\ref{tab:energy}. Released turbulent energies are comparable to each other for both transients. Interestingly, we noticed that amount of turbulent energy released in cool transient is an order of magnitude larger than other thermal energies released whereas all the released energies are comparable in the hot transient. \citet{2018MNRAS.479.2382L} had also studied small-scale transients of temperature 1.85 MK and found that amount of turbulent energy (obtained from broadening of Si IV 1394 \AA\ line profiles) is an order of magnitude larger than involved thermal energies. These are interesting findings and demands detailed statistical study on any relation between temperature of plasma achieved and amount of thermal and turbulent energies released.

Recent MHD simulation results of \citet{2016ApJ...832..195N} suggests that plasma$-\beta$ is an important factor to decide on temperature of plasma that can be heated up. They speculated that maximum temperature achieved will increase with lowering of plasma$-\beta$. \citet{2019A&A...628A...8P} also investigated effect of plasma$-\beta$ on reconnection and found that under low plasma$-\beta$, temperature and velocities in the reconnection region will be increasing with decreasing plasma$-\beta$. Therefore, we utilize the results obtained from magnetic field extrapolation and obtained density and temperature of loop segments to test their findings, if possible. We selected similar volume element as used in density measurements from DEM analysis to get the estimates on coronal magnetic fields. On an average, we found field strength to be about 143 G above the transient location along the hot loop segment at the height of 0.72 Mm with respect to photospheric magnetic field measurement. Similarly, field strengths of about 62 G, 25 G, and 55 G were found along the different loop segments of cool loops at the heights of about 0.18 Mm, 0.75 Mm, and 1.08 Mm respectively. First two cool loop segments are smaller in length whereas last one is bigger in length and reaches high in the atmosphere. However irrespective of length of these loop segments, our electron number density estimates obtained from AIA DEM analysis are of the order of $8\times 10^{9}$ cm$^{-3}$ and $2.5 \times 10^{9}$ cm$^{-3}$ and are almost constant along the hot and cool loops respectively. Using these values, estimates on plasma-$\beta$ is of the order of $10^{-5}-10^{-4}$ for both loops. Given the error-bars on magnetic field measurements and extrapolated fields, inferring any difference in plasma-$\beta$ of hot and cool loops is almost impossible. Therefore, no conclusion can be drawn in this study on any influence of plasma-$\beta$ on plasma temperature achieved by these transients and associated hot and cool loops.

Transient hot and cool loops can be cooled down to its background atmospheric temperature via electron thermal conduction and radiative losses. Lifetime  of observed loops at different temperatures and densities are given by $\tau \approx 2.35 \times 10^{-2} {L_0^{5/6}}/{T_e^{1/6}/n_e^{1/6}}$ due to the combined cooling effect of conduction and radiation \citep[e.g.,][]{1995ApJ...439.1034C}. Therefore, we took half loop length $L_0\approx 10$ Mm, temperature and number density for hot component of hot loop and found $\tau_{hloop}^{hot} \approx 19$ min. Similarly upon using all the parameters as mentioned in Table~\ref{tab:summary}, we obtained $\tau_{hloop}^{cool} \approx 23$ min,  $\tau_{cloop}^{cool} \approx 38$ min, and $\tau_{cloop}^{hot} \approx 55$ min. Longer life time of both hot and cool loops obtained here is consequences of our assumption of treating these loops as a single long loops. As noted earlier, these loops are collection of several small loops (shorter loop lengths) which also have fine-threads within. Thus, if smaller loop lengths are assumed for these loops then obtained life time of these loops can easily match with observed life times. Cool loop will have even much shorter loop lengths as compared to hot loop. These small loop segments are quite well resolved in the magnetic field extrapolations (see  Fig.~\ref{fig:extrapolated}), and thus supports our claim on small length of loop segments. These findings of small segments of loops within the cool loop system adds further complications in our understanding of formation of cool loop system.

In summary, we studied small-scale transients which led to the formation of transient hot and cool loops. These transients and loops were multi-thermal in nature and attained maximum temperature of about 8 MK and 0.4 MK respectively. These events were driven by magnetic reconnections as evident from flux cancellations happening underneath. Spectroscopic study revealed that not only thermal energies but significant amount of turbulent energies were also released during the transients and was order of magnitude larger fro cool transient. Life time estimates and magnetic field extrapolation suggested the presence of small-scale and fine structures within the loops. Energy estimates indicate that flux cancellation events can easily power the hot transient but can not explain that for cool transients. This could be due to poor sensitivity and spatial resolution of HMI magnetogram. Thus, future studies with high resolution magnetic maps obtained from DKIST \citep{2020SoPh..295..172R}, MAST \citep{2017CSci..113..686V}, and PHI/SO \citep{2020A&A...642A..11S} will shed more light on such small-scale transients and formation of hot and cool loops. Furthermore, to get more insight on conditions leading to formation of hot and cool loops in the solar atmosphere, a statistical study combined with modeling is needed to get detailed account on electron density, temperature, magnetic flux decay, thermal and turbulent energies released and plasma-$\beta$ of both such loops to get the complete picture and understanding of physics of such events. This will also enable us to get comprehensive understanding on the existence of hot plasma in the core of active regions \citep[e.g.][]{2021ApJ...909..105T}.

\section*{Acknowledgments}
We thank the referee for the comments that improved quality of manuscript considerably. AIA and HMI data are courtesy of SDO (NASA). IRIS is a NASA small explorer mission developed and operated by LMSAL with mission operations executed at NASA Ames Research center and major contributions to downlink communications funded by the Norwegian Space Center (NSC, Norway) through an ESA PRODEX contract. Facilities: SDO (AIA, HMI). CHIANTI is a collaborative project involving George Mason University, the University of Michigan (USA) and the University of Cambridge (UK).

\section*{Data Availability}

The observational data utilized in this study from AIA and HMI on-board SDO are available at   \url{http://jsoc.stanford.edu/ajax/lookdata.html} and data from IRIS mission are available at https://www.lmsal.com/hek/hcr?cmd=view-recent-events\&instrument=iris. Atomic data used for spectroscopic diagnostics are available at \url{https://www.chiantidatabase.org} and distributed via Solarsoft \url{https://www.lmsal.com/solarsoft/}.







\bsp	

\label{lastpage}

\end{document}